\documentclass[11pt, oneside]{article}   	
\usepackage{geometry}                		
\usepackage{graphicx}				
\usepackage{amssymb}
\usepackage{amsmath}

\usepackage{amsthm}

\title{ Algorithm for SIS and MultiSIS problems}
\author{Igor Semaev\\ Department of Informatics, University of Bergen, igor@ii.uib.no}

\begin{document}
\maketitle
\begin{abstract}
 SIS problem has numerous applications in cryptography. Known algorithms for solving that problem are exponential in complexity. A new algorithm is suggested in this note, its complexity is sub-exponential for a range of parameters.
\end{abstract}

\section{Introduction}\label{intro}
Let $A$ be  any integer $m\times n$ matrix, where $m>n$ and $q$ be a prime. Assume $A$ is of rank $n$ modulo $q$.  Let $c=(c_1,\ldots,c_m)$ be an integer vector of length $m$ and $|c|=(c_1^2+\ldots+c_m^2)^{1/2}$ denote  its norm (Euclidean length) and
 $\nu$ be a positive real. The SIS (Short Integer Solution) problem is to construct a non-zero integer row vector $c$ of length $m$ and norm at most $\nu$ such that $cA\equiv 0\mod q$. The problem of constructing  several such short vectors is called MultiSIS problem.  
 
  The inhomogeneous SIS problem asks for a short vector $c$ such  that $cA\equiv a\mod q$ for a non-zero row vector $a$ of length $n$. The inhomogeneous SIS problem may be reduced to a homogeneous SIS problem. Let $A_1=\begin{pmatrix} A\\ a\end{pmatrix}$ be a concatenation of the matrix $A$ and the  vector $a$.   Assume one constructs a number of short solutions $c_1$ to $c_1A_1\equiv 0\mod q$ with non-zero last entry.  One of them may likely be $c_1=(c,1)$ and that gives a solution to $cA\equiv a\mod q$, or such a vector may be found as a combination of the solutions to the SIS problem.   
 
 Typical SIS problem parameters   are $\nu\geq\sqrt{n\log_2q}$ and $m> n\log_2q$, where $q$ is bounded by a polynomial in $n$. 
 The problem may be reduced to constructing short vectors in general lattices, which is considered hard, see \cite{A}. The SIS problem has a number of applications in cryptography, see \cite{P15}. For instance,  the hash function $x\rightarrow xA$ was  suggested in \cite{A}.    
 
  Integer vectors $c$ such that $cA\equiv 0\mod q$  is a  lattice of dimension $m$ and volume $q^n$. So all  vectors of norm $\leq \nu$  may be computed with a lattice enumeration in time  $m^{O(m)}$, see \cite{HPS}. Alternatively, one may  apply a lattice reduction algorithm. The  reduction    cost  is $2^{O(m)}$ operations according to \cite{HPS}. The so-called combinatorial algorithms to solve the SIS problem and its inhomogeneous variant, where the entries of $c$ are $0$ or $1$,     are surveyed in  \cite{BGL19}. They  have complexity $2^{O(m)}$ operations.  All above methods are thus exponential in complexity. In this note a new algorithm for solving SIS and MultiSIS problems is introduced. The complexity of the algorithm is sub-exponential for a range of parameters.    

\section{MultiSIS Problem}\label{low_norm}

  How to construct $N$ different non-zero vectors $c$ of norm at most $ \nu$  such that $cA\equiv 0\mod q$? The vectors generated by the rows of the matrix $qI_m$, where $I_m$ denotes a unity matrix of size $m\times m$, are trivial solutions and not counted.  We call this MultiSIS problem. Obviously, a solution to the MultiSIS problem implies a solution to the homogeneous SIS problem. That may also imply a solution to a relevant inhomogeneous problem as it is explained  earlier.
 
 The MultiSIS problem may be solved by lattice enumeration. Alternatively, one perturbs the initial basis of the lattice $N$ times and apply a lattice reduction algorithm after each perturbation.  So the overall complexity is $N2^{O(m)}$, though we do not know if that really  solves the problem as the vectors in the reduced bases may repeat. 
 
 If $m=o(\nu^2)$, then the number of integer vectors $c$ of norm at most $\nu$ is approximately the volume of a ball of radius $\nu$ centred at the origin. With probability $1/q^n$ the vector $c$ satisfies $cA\equiv 0$. Therefore the number of such relations is around 
 $$\frac{\pi^{m/2}\,\nu^m}{\Gamma(m/2+1)\,q^n}\approx \frac{(2\pi e)^{m/2}}{\sqrt{\pi m}}\left(\frac{\nu}{\sqrt{m}}\right)^m\frac{1}{q^n}$$ 
and should be at least $N$ to make the problem solvable. That fits the so-called Gaussian heuristic, see \cite{LLL_book}.

 According to \cite{MO},  if $\nu=O(\sqrt{m})$ the Gaussian heuristic does not generally hold. We will use a different argument  still heuristic. 
Let $ \nu<\sqrt{m}$ and $d=\lfloor \nu^2\rfloor$. For each subset $A_{i_1},\ldots,A_{i_r}$ of $r\leq d$ rows of $A$ there are $2^r$ linear combinations $c_1A_{i_1}+\ldots +c_rA_{i_r}$, where $c_i=\pm 1$ and so $c=(c_1,\ldots,c_r)$ is of norm  $\leq \nu$. We do not distinguish between $c$ and $-c$. So the expected number of such zero  combinations is  $2^{r-1}/q^{n}$. For the whole matrix  the expected number of different  $c$  of norm at most $\nu$ such that $cA\equiv 0$ is at least $\sum_{r=1}^d {m\choose r}2^{r-1}/q^{n}$. 
Therefore, $N$ such relations do exist if  $\sum_{r=1}^d {m\choose r}2^{r-1}/q^{n}\geq N$, minding that the inequality is approximate.

\subsection{MultiSIS Algorithm}
Let $\delta=m/n\ln q$ and $\eta=\nu^2/n\ln q$.
 In this section we present the algorithm to construct vectors $c$ of norm at most $\nu$ such that $cA\equiv 0\mod q$. In Section \ref{analysis} we will  show that if at least one of $\delta$ or $\eta$ tends to infinity, then one may construct $q^{\frac{n}{t}(1+o(1))}$ such vectors with  the  complexity  $q^{\frac{n}{t}(1+o(1))}$operations, where $t=[\log_2\sqrt{\eta\ln\delta}]\,(1+o(1))$. The latter    tends to infinity, so the complexity is  sub-exponential. If both $\delta$ and $\eta$ are bounded, then the complexity is represented by the same expression   for some bounded $t$ and therefore exponential.  The  analysis is heuristic. 
 
 Let $d\geq 2,k<m,N$  be integer parameters such that $\nu=d\sqrt{k}$. We may assume that $d=2^t$ for an integer $t=\log_2d$ and $n=st$ for an integer $s$. Otherwise, the algorithm below is easy to adjust. Let $\mathfrak{m}(k)$ be the number of integer vectors of length $m$ and of norm $\leq \sqrt{k}$ up to a multiplier $-1$. It is easy to see that $\mathfrak{m}(k)\geq \sum_{i=1}^k {m\choose i}2^{i-1}$.

\begin{enumerate}
\item  
Put   $\mathfrak{A}_0=C_0A$, where
 $C_0$ be a matrix of size $\mathfrak{m}(k)\times m$ and each row of $C_0$ is an integer vector of norm at most  $\sqrt{k}$.

\item  Let $N_i$ for  $i$ in $0,\ldots,t-1$ be integers such that $N_i= q^{s(1+o(1))}$, where 
 $N_0\leq\mathfrak{m}$ and $N_t=N$.

 \item For $i=0,\ldots,t-1$ do the following. Represent
$\mathfrak{A}_i=\mathfrak{A}_{i1}|\mathfrak{A}_{i2}$ as a concatenation of two matrices, where  $\mathfrak{A}_{i1}$ is of size $N_i\times s$ and $\mathfrak{A}_{i2}$ is of size $N_i\times s(t-i-1)$. As $N_i= q^{s(1+o(1))}$     there are  $ N_{i+1}= q^{s(1+o(1))}$ relations $c\,\mathfrak{A}_{i1}\equiv0$, where
$c$ is a vector of length $N_i$ and it has  at most two non-zero  entries which are $\pm 1$. Let
 $C_{i+1}$ be a matrix of size $N_{i+1}\times N_{i}$ with such rows.  Equivalently, there are $q^{s(1+o(1))}$ pairs of  rows in $\mathfrak{A}_{i1}$, where one row differs from another   by a multiplier $\pm 1$,   and zero rows in $\mathfrak{A}_{i1}$.   
Such pairs of rows and zero rows in $\mathfrak{A}_{i1}$ may be computed in $N_i^{1+o(1)}$ operations by sorting. 
 Put $\mathfrak{A}_{i+1}=C_{i+1}\mathfrak{A}_{i2}$ and repeat the step.
 \item The matrix  $C=C_t\ldots C_1C_0$ is of size $N\times m$ and it satisfies $CA\equiv 0$. Each row of $C$ has norm $\leq \nu=d\sqrt{k}$. 
\end{enumerate}

The rows of $C_0$ are different and non-zero. 
 At each step of the algorithm one may choose $C_{i}$ such that the rows of $C_i\ldots C_1 C_0$ are different. As the rows of $C_{i+1}$ have at most two non-zero  entries which are $\pm 1$, the rows of $C_{i+1}C_i\ldots C_0$ are all non-zero. Though we can not guarantee theoretically that all constructed vectors are different, the algorithm works well in practice.
 
\subsection{Analysis of the Algorithm}\label{analysis} 
The algorithm constructs $q^{\frac{n}{t}(1+o(1))}$ integer vectors $c$ of norm at most $\nu$ such that $cA\equiv 0\mod q$ and its  complexity is $q^{\frac{n}{t}(1+o(1))}$ operations. We  will define an optimal   $t=\log_2d$. For any input parameters $n,q,m,\nu$ one may find $t$ by solving numerically the system $\mathfrak{m}(k)\geq q^{\frac{n}{t}}$ and $\nu=2^t\sqrt{k}$.

Let $\delta=m/n\ln q$ and $\eta=\nu^2/n\ln q$ and at least one of them is an increasing function in $n$. We will represent $t$ as a function of $\delta,\eta$.
First, we find $k$ such that
$\mathfrak{m}(k)\geq q^{\frac{n}{t}}$ for large $n$. One may solve a stronger inequality ${m\choose k}2^{k-1}\geq q^{\frac{n}{t}}$ instead. With the Stirling approximation to the factorial function, it is easy to see that one may take 
$k= \frac{\alpha n}{t}(1+o(1))$, where 
$$\alpha=\frac{\ln q}{\ln m-\ln\ln q^{\frac{n}{t}} }=\frac{\ln q}{\ln(\delta t)}.
$$
So $k=\frac{n\ln q}{t\ln(\delta t)}(1+o(1))$ and the equation $\nu=d\sqrt{k}$ is equivalent to
\begin{equation}\label{delta_eta}
\eta=\frac{4^t}{t\ln (\delta t)}(1+o(1)).
\end{equation}
 The solution to \eqref{delta_eta} is 
$$t=\log_2\sqrt{\eta\ln\delta}\,(1+o(1)).$$
Experimentally, $t>\log_2\sqrt{\eta\ln\delta}$ and they converges for very large parameters.
The complexity of the algorithm is $q^{\frac{n}{\log_2\sqrt{\eta\ln\delta}}(1+o(1))}$.

\bibliographystyle{alpha}

\end{document}